\documentclass[journal]{IEEEtran}
\usepackage{amsfonts}
\usepackage{cite}
\usepackage[pdftex]{graphicx}
\usepackage{amsmath}
\usepackage{amssymb}
\usepackage{bbm}
\usepackage{mathtools}
\usepackage{suffix}
\usepackage{array}
\usepackage[cal=zapfc]{mathalfa}
\usepackage{adjustbox}
\usepackage{url}
\usepackage{xcolor}
\usepackage{stfloats}
\usepackage{bigfoot}
\usepackage{bm}
\usepackage{soul}
\usepackage{multirow}
\usepackage{epstopdf}
\usepackage{booktabs}
 
\usepackage[caption=false,font=footnotesize]{subfig}

\usepackage{tikz}
\usetikzlibrary{tikzmark}
\usetikzlibrary {arrows.meta}

\ifCLASSINFOpdf
\DeclareGraphicsExtensions{.pdf,.jpeg,.png}
\else
\DeclareGraphicsExtensions{.eps}
\fi
\def \am {$\alpha$-$\mu$}
\def \kap {$\kappa$-$\mu$}
\def \km {$\kappa$-$\mu$}
\def \et {$\eta$-$\mu$}

\def \km {$\kappa$-$\mu$}
\def \naka {Nakagami-\emph{m}}

\DeclarePairedDelimiterX\MeijerM[3]{\lparen}{\rparen}
{\,#1\:\delimsize\vert \: \begin{matrix}#2 \\[0.5em] #3 \end{matrix}\,}
\newcommand\MeijerG[8][]{  G^{#2,#3}_{#4,#5}\MeijerM[#1]{#6}{#7}{#8}}
\WithSuffix
\newcommand\MeijerG*
[7]{
  G^{#1,#2}_{#3,#4}\MeijerM*{#5}{#6}{#7}}

\DeclarePairedDelimiterX\FoxM[3]{\lparen}{\rparen}
{\,#1\:\delimsize\vert \: \begin{matrix}#2 \\[0.5em] #3 \end{matrix}\,}
\newcommand\FoxH[8][]{  H^{#2,#3}_{#4,#5}\FoxM[#1]{#6}{#7}{#8}}
\WithSuffix
\newcommand\FoxH*
[7]{
  H^{#1,#2}_{#3,#4}\FoxM*{#5}{#6}{#7}}  
  
  \usepackage{pdfpages}
  
  \usepackage{mathtools}

\begin{document}

 \title{Cute but Cunning: Effective Closed-Form Alternatives to the Exact Lognormal Statistics}
\author{Carlos~Rafael~Nogueira~da~Silva,~Maria~Cecilia~Luna~Alvarado,~Fernando~Darío~Almeida~García,~\IEEEmembership{Senior Member,~IEEE},~and~Michel~Daoud~Yacoub,~%
\IEEEmembership{Member,~IEEE}
\thanks{This work was supported by CNPq (Grant Reference: 301.581/2019-3 and 420.148/2023-0), and by FAPESP (Grant Reference: 2023/02578-1). This work was also funded by the Brasil 6G Project with support from RNP/MCTI (Grant 01245.010604/2020-14), and the xGMobile Project code XGM-AFCCT-2025-8-1-1 with resources from EMBRAPII/MCTI (Grant 052/2023 PPI IoT/Manufatura 4.0) and FAPEMIG Grant PPE-00124-23.}%
\thanks{C.\ R.\ N.\ da Silva is with the Department of Electrical Engineering, Federal University of Triângulo Mineiro, Uberaba 38064-200, Brazil (e-mail: $ \rm carlos.nogueira@ \rm uftm.edu.br$).}%
\thanks{M.\ D.\ Yacoub and M.\ C.\ L. Alvarado are with the Wireless Technology Laboratory, School of Electrical and Computer Engineering, University of Campinas, Campinas 13083-970, Brazil (e-mail: $ \rm  mdyacoub@\rm unicamp.br$, $ \rm  mcla207@\rm gmail.com)$.}
\thanks{F. D. A. García is with the Wireless and Artificial Intelligence (WAI) laboratory, National Institute of Telecommunications (INATEL), Santa Rita do Sapucaí, MG, 37540-000, Brazil (e-mail: \mbox{fernando.almeida@inatel.br}).}
}

\maketitle
\begin{abstract}
The Lognormal distribution is a fundamental statistical model widely used in different fields of science, including biology, finance, economics, engineering, etc. In wireless communications, it is the primary statistic for large-scale fading modeling. However, its known analytical intractability presents persistent channel characterization and performance analysis challenges. This paper introduces two effective and mathematically tractable surrogate models for the Lognormal distribution, constructed from the product of Nakagami-$m$ and Inverse Nakagami-$m$ (I-Nakagami-$m$) variates. These models yield asymptotically exact closed-form expressions for key performance metrics---including the characteristic function, bit error rate, and Shannon’s capacity---and enable analytically tractable expressions for the probability density function and cumulative distribution function of the composite $\alpha$-$\mu$-Lognormal fading model.
To facilitate implementation, a moment-matching framework is developed to map the Lognormal parameters to the surrogate model parameters. In addition, a random mixture approach is proposed to enhance convergence by exploiting the complementary approximation properties of the Nakagami-$m$ and I-Nakagami-$m$ distributions. The methodology is further extended to heterogeneous cascaded fading channels comprising arbitrary combinations of $\alpha$-$\mu$, $\kappa$-$\mu$, and $\eta$-$\mu$ variates, for which moment-based mappings to the equivalent Lognormal distributions are derived. 
Numerical results confirm the accuracy and efficiency of the proposed approach, positioning it as a practical and reliable alternative to exact Lognormal statistics.
\end{abstract}


\section{Introduction}


The Lognormal distribution stands as one of the most fundamental statistical models across numerous scientific disciplines, including biology, finance, economics, engineering, etc. In wireless communications, it is the primary and most fundamental choice to model large-scale fading. Additionally, it has been employed to approximate the distribution of product of random variates (e.g., \cite{Zheng2012}) which arises in cascaded channels or multihop systems. Despite its widespread applicability, performing analytical operations under this distribution often proves exceedingly complex or infeasible. This complexity is usually circumvented by the  adoption of alternative, approximate, models as substitutes for the Lognormal statistics~(e.g.,\cite{Yoo:2017,9380972,Abdi:1999,ip-com_20045126
}), namely, the probability density function (PDF) and the cumulative distribution function (CDF).
For instance, in~\cite{Yoo:2017} and \cite{9380972}, extensive empirical validation supported by field measurements and statistical tests demonstrated that the Inverse Gamma (InvG) distribution can accurately model large-scale fading, making it a suitable alternative to the Lognormal distribution. Furthermore, in~\cite{Yang2025}, a comprehensive information-theoretic analysis based on the Kullback–Leibler (KL) divergence was conducted to investigate the feasibility, mathematical tractability, and optimality of employing the InvG distribution as a substitute for the Lognormal model. In~\cite{Abdi:1999,ip-com_20045126}, the Gamma distribution was proposed as another alternative to the Lognormal distribution. In~\cite{Dang2023}, the distance between the original Lognormal and the approximating Gamma distributions was quantified using the Kullback–Leibler (KL) divergence, whereas~\cite{Wang2024} employed the Kolmogorov–Smirnov criterion to derive statistically robust parameter-mapping relations between the two distributions. 
While the above substitutes may retain certain statistical features of the original distribution, they often fall short in accurately capturing its full behavior. 
Although they facilitate mathematical analysis by enabling closed-form expressions, they inherently involve a trade-off between precision and tractability. This fundamental compromise poses a significant challenge for researchers and practitioners who depend on the Lognormal distribution for correct wireless modeling and performance analysis.


Lognormal-based fading models play a central role in accurately characterizing wireless propagation environments where composite large-scale (shadowing) and small-scale (multipath) fading coexist. Notable examples include the Nakagami-Lognormal\cite{Tjhung99}, Weibull-Lognormal\cite{Karadimas08,Karadimas09}, Rayleigh-Lognormal (Suzuki)\cite{Suzuki77}, Rice-Lognormal\cite{Vatalaro95}, $\alpha$-$\eta$-$\mu$-Lognormal and $\alpha$-$\kappa$-$\mu$-Lognormal\cite{Bhatt18} distributions, each capturing distinct physical mechanisms of multipath and shadowing effects.
Despite their widespread relevance in wireless modeling, a major challenge in analyzing these composite distributions lies in the absence of either closed-form or tractable expressions for the key statistical functions, such as the PDF and CDF. This limitation stems primarily from the analytical intractability of the Lognormal distribution itself, which prevents the exact evaluation of the integrals that arise in composite channel modeling.
Consequently, direct analysis of such models is mathematically challenging. To address this, researchers often resort to numerical integration, approximate approaches, or Monte Carlo simulation techniques---methods that are computationally intensive and provide limited analytical insight\cite{Bhatt18,Hansen77,Hussaini02,Corazza94,Abdi98,Singh17}. 
These challenges motivate the development of alternative modeling strategies that retain the core statistical characteristics of Lognormal-based fading, while enabling closed-form solutions and improved analytical tractability for performance analysis and system design.

Given the limitations of existing methods in handling the Lognormal-based analyses, there is a clear and pressing need to develop more robust solutions that maintain the essential statistical features of the Lognormal distribution while enabling tractable mathematical treatment. 
An ideal approach would preserve the key statistical properties that underpin the Lognormal distribution's broad applicability, while overcoming the analytical hurdles that have long limited its direct application. 
In our previous work \cite{10883648}, we developed an exact replacement for the Lognormal PDF using the product of random variables following Power and/or Pareto distributions. 
This formulation departs from traditional approximation methods by offering a mathematically equivalent representation that preserves the Lognormal’s core statistical structure while significantly improving analytical tractability. 
By leveraging the multiplicative properties of the Power and Pareto distributions, this advancement enables more precise analysis in complex scenarios where traditional methods prove inadequate, potentially easing the way Lognormal distributions are handled across various scientific and engineering applications. 
However, as explicitly acknowledged in~\cite{10883648}, the use of Power and Pareto distributions was not intended as the final or optimal choice. Rather, it was recognized as a foundational step, with the possibility that alternative surrogate models could offer better convergence properties or simplified formulations. 
Motivated by this, we propose an extension of this methodology by introducing two additional surrogate configurations: one based on the product of Nakagami-$m$ variates and another based on Inverse Nakagami-$m$ variates, denoted here as I-Nakagami-$m$. These new formulations offer improved flexibility for modeling complex random processes and provide alternative analytical perspectives for dealing with the Lognormal distribution. By exploring these product-based models, we aim to further advance the analytical treatment of Lognormal and Lognormal-based fading distributions---bridging the longstanding gap between statistical accuracy (fidelity) and mathematical tractability.


\subsection{Contributions}

The contributions of this work are summarized as follows:

\begin{enumerate}
    \item We introduce two alternative surrogate models for the Lognormal distribution, constructed from the product of Nakagami-$m$ and I-Nakagami-$m$ variates. These models yield analytically tractable, closed-form expressions and exhibit faster asymptotic convergence to the Lognormal distribution compared to previous surrogate models, namely, the Pareto and Power distributions.
    
    \item To support practical implementation, we establish a bidirectional mapping between the Lognormal distribution and its Nakagami-$m$ and I-Nakagami-$m$ surrogates. Specifically, we derive closed-form expressions that relate the Lognormal parameters to their Nakagami-$m$ and I-Nakagami-$m$ counterparts using logarithmic moments. 
    
    \item Leveraging the analytical tractability of the proposed surrogate models, we derive asymptotically exact expressions for several key performance metrics in a wireless system affected by Lognormal fading, including the characteristic function (CF), bit error rate (BER), and Shannon’s capacity. Additionally, these models enable asymptotically exact solutions for the PDF and CDF of the composite $\alpha$-$\mu$-Lognormal fading model.\footnote{The $\alpha$-$\mu$-Lognormal fading model can be obtained from the $\alpha$-$\kappa$-$\mu$-Lognormal distribution by setting $\kappa=0$\cite{Bhatt18}.}
    
    \item We propose a convergence enhancement strategy that blends the Nakagami-$m$ and I-Nakagami-$m$ distributions through a probabilistic (random) mixture. This hybrid model substantially reduces the number of product terms required to attain a desired level of precision.

    \item We extend the analysis to heterogeneous cascaded fading models (i.e., products of random variables governed by different statistical distributions) encompassing arbitrary combinations of $\alpha$-$\mu$, $\kappa$-$\mu$, and $\eta$-$\mu$ variates. 
    For these composite fading scenarios, we derive explicit moment-based mappings that connect the parameters of the resulting product distributions to those of an equivalent Lognormal model, enabling a unified and tractable statistical representation.
    
\end{enumerate}

\subsection{Structure and Notation}

The remainder of this paper is organized as follows. Section~\ref{sec: Preliminaries} presents the necessary background on fading models, the definition of the mean and instantaneous SNR, and the connection between product distributions and the Lognormal model. Section~\ref{sec: Nakagami and Lognormal Mapping} introduces the Nakagami-$m$ product-based surrogate, derives its Lognormal mapping, and presents closed-form expressions for key performance metrics. Section~\ref{sec: I-Nakagami and Lognormal Mapping} develops an alternative surrogate based on the product of I-Nakagami-$m$ variates and provides corresponding expressions. Section~\ref{sec: Convergence Improvement} proposes a convergence enhancement strategy using a probabilistic mixture of Nakagami-$m$ and I-Nakagami-$m$ products. Section~\ref{sec:usefull} extends the analysis to heterogeneous cascaded fading models involving $\alpha$–$\mu$, $\kappa$–$\mu$, and $\eta$–$\mu$ distributions. Section~\ref{sec: Some Plots} presents numerical results that validate the proposed models. Final remarks are given in Section~\ref{sec: Final Remarks}.

In the sequel, $\mathbb{E}[\cdot]$ denotes expectation; $\mathbb{V}[\cdot]$, variance; $\text{sign}(\cdot)$, the sign function; $\Gamma(\cdot)$, the gamma function~\cite[Eq. (6.1.1)]{Abramowitz:1972}; $\Gamma(\cdot, \cdot)$, the upper incomplete Gamma function~\cite[Eq. (6.5.3)]{Abramowitz:1972}; $\psi^{(\cdot)}(\cdot)$, the polygamma function \cite[Eq. (6.4.1)]{Abramowitz:1972}; $I_n(\cdot)$, the modified Bessel function of the first kind and order $n$ \cite[Eq. (9.6.10)]{Abramowitz:1972}; $(\cdot)_{(\cdot)}$, the Pochhammer's symbol \cite[Eq. (6.1.22)]{Abramowitz:1972}; ${}_1F_1(\cdot,\cdot,\cdot)$, the Kummer's confluent hypergeometric function \cite[Eq. (13.1.2)]{Abramowitz:1972}; $\, _2F_1\left(\cdot,\cdot;\cdot;\cdot\right)$, the Gauss hypergeometric function~\cite[Eq. (15.1.1)]{Abramowitz:1972}; $G^{a,b}_{c,d} \left[\cdot \right]$, the Meijer-G function \cite[Eq. (8.2.1.1)]{Prudnikov:1986c}; $H^{a,b}_{c,d} \left[\cdot \right]$, the Fox-H function\cite[Eq. (1.2)]{Mathai:2009}; and $j=\sqrt{-1}$, the imaginary unit.

\section{Preliminaries}
\label{sec: Preliminaries}

This section provides a brief review of key fading models, including the Lognormal, $\alpha$-$\mu$, $\kappa$-$\mu$, and $\eta$-$\mu$ distributions. It also introduces the definition of the mean and instantaneous signal-to-noise ratio (SNR), and outlines the connection between product distributions and the Lognormal model, which forms the basis for the surrogate approaches proposed in this work.

\subsection{Fading Models of Interest}

\subsubsection{The Lognormal Distribution}
Let $R > 0$ be a random envelope lognormally distributed with parameters $\nu = \mathbb{E}[\log(R)]$ and $\sigma^2 = \mathbb{V}[\log(R)]$. The PDF of $R$ is given by
\begin{equation}
    \label{eq:pdf_lognormal}
    f_R(r)=\frac{1}{\sqrt{2 \pi } r \sigma } \exp\left({-\frac{(\log (r)-\nu
   )^2}{2 \sigma ^2}}\right).
\end{equation}
The higher-order moment are given as
\begin{equation}\label{eq:log_moment}
    \mathbb{E}[R^k] = e^{k \nu +\frac{k^2 \sigma ^2}{2}}.
\end{equation}
In wireless communications, this is the fundamental model that captures the effects of long-term fading\footnote{Of course, some alternative models, such as Nakagami-$m$, I-Nakagami-$m$, and others, assumed to yield reasonable approximations, are used instead to circumvent the mathematical intricacy of the Lognormal distribution.}. 

\subsubsection{The \am{} Fading Distribution}
Let $R > 0$ be a fading signal that follows the \am{} fading model with  $\hat{r}^\alpha = \mathbb{E}[R^\alpha]$. The PDF of $R$ is given by
\begin{equation}
    \label{eq:pdf_au}
    f_R(r) = \frac{\alpha \mu^{\mu}}{\Gamma(\mu)} \frac{r^{\alpha \mu - 1}}{\hat{r}^{\alpha \mu }} \exp\left({- \mu\frac{r^\alpha}{\hat{r}^\alpha}}\right),
\end{equation}
in which $\mu>0$ is related to the number of multipath cluster and $\alpha>0$ is the non-linearity coefficient.
The higher-order moments are obtained as
\begin{equation}\label{eq:au_moment}
    \mathbb{E}[R^k]=\frac{\Gamma \left(\frac{k}{\alpha }+\mu \right)}{\Gamma (\mu )}
   \left(\frac{\hat{r}}{\mu ^{1/\alpha }}\right)^k.
\end{equation}
In addition to the higher-order moments, the first and second logarithmic moments might be useful, and, for the \am{} distribution, they are obtained as
\begin{equation}\label{eq:au_logmean}
\nu_\alpha=\mathbb{E}[\log(R)] = \log \left(\frac{\hat{r}}{\mu ^{1/\alpha }}\right)+\frac{\psi
   ^{(0)}(\mu )}{\alpha },
\end{equation}
and
\begin{equation}
    \sigma_\alpha^2=\mathbb{V}[\log(R)] = \mathbb{E}[\log^2(R)]-\mathbb{E}[\log(R)]^2 = \frac{\psi ^{(1)}(\mu )}{\alpha ^2}.
\end{equation}

The \am{} fading model has gained a lot of attention recently as a prominent candidate to model terahertz signal. It is a generalization of the \mbox{\naka{}} fading model that includes, as special cases, the Rayleigh, \mbox{\naka{}}, Negative Exponential, Semi-Normal, and Weibull. 

\subsubsection{The \kap{} Fading Model}
Let $R> 0 $ be a fading signal that follows a \kap{} distribution with rms value $\hat{r}_i^2 =\mathbb{E}[R^2]$. The PDF of $R$ is 
\begin{equation}
\begin{split}
    f_R(r)&= \frac{2 \mu_i  (1+\kappa_i)^{\frac{\mu_i +1}{2}}}{\kappa_i ^{\frac{\mu_i -1}{2}} \exp (\kappa_i  \mu_i)} \frac{r^{\mu_i}}{\hat{r}_i^{\mu_i +1}} \exp
   \left(-\frac{\mu_i  (1+\kappa_i ) r^2}{\hat{r}_i^2}\right)\\&\times I_{\mu_i -1}\left(\frac{2 \mu_i  r \sqrt{\kappa_i  (1+\kappa_i )}}{\hat{r}_i}\right),
\end{split}
\end{equation}
in which $\kappa_i$ is the ratio between the total power of the dominant components and the total power of the scattered waves, and $\mu_i>0$ is related to the number of multipath clusters.
The higher-order moments are obtained as
\begin{equation}\label{eq:ku_moment}
    \begin{split}
        \mathbb{E}[R^K] = \frac{\hat{r}_i^k (\mu_i )_{\frac{k}{2}}}{((1+\kappa_i ) \mu_i )^{k/2}} \, _1F_1\left(-\frac{k}{2};\mu_i ;-\kappa_i  \mu_i \right).
    \end{split}
\end{equation}
Its logarithmic mean is found as
\begin{equation}
\begin{split}\label{eq:ku_logmean}
   \nu_\kappa^{(i)}= \mathbb{E}[\log(R)] &= \log \left(\frac{\hat{r}_i}{\sqrt{(1+\kappa_i ) \mu_i }}\right)+\frac{\psi ^{(0)}(\mu_i)}{2}\\
    &-\frac{1}{2} {}_1{F_1}^{(1,0,0)}(0,\mu_i ,-\kappa_i  \mu_i ),
\end{split}
\end{equation}
in which ${}_1{F_1}^{(1,0,0)}(a,b,z)$ is the first derivative of the Kummer’s confluent hypergeometric function with respect to $a$ \cite[Eq. (07.20.20.0001.01)]{wolfram}.
Although a closed-form expression for the logarithmic variance could not be obtained, it can be efficiently computed through numerical methods.

\subsubsection{The \et{} Fading Model}
Let $R> 0 $ be a fading signal that follows the \et{} distribution with rms value $\hat{r}^2 = \mathbb{E}[R^2]$. The PDF of $R$ given by
\begin{equation}
\begin{split}
        f_R(r) &= \frac{4 \sqrt{\pi } \mu ^{\mu +\frac{1}{2}} h^{\mu }}{\Gamma (\mu ) H^{\mu -\frac{1}{2}}} \frac{r^{2 \mu }}{\hat{r}^{2 \mu +1}} \exp
   \left(-\frac{2 \mu  h r^2}{\hat{r}^2}\right)\\&\times I_{\mu -\frac{1}{2}}\left(\frac{2 \mu  H r^2}{\hat{r}^2}\right),
\end{split}
\end{equation}
in which $\mu>0$ is related to the number of multipath cluster, $h$ and $H$ depend on a parameter $\eta$ and are defined in accordance to the chosen \et{} format. In Format 1, the parameter $\eta >0$ is defined as the ratio between the power of the in-phase and the power of the quadrature waves and $h = (2+\eta^{-1}+\eta) / 4$ and $H = (\eta^{-1}-\eta) / 4$. In format 2, the parameter $-1<\eta <1$ is the correlation coefficient between the in-phase and quadrature waves and $h = 1 / (1-\eta^2)$ and $H = \eta h$. The \et{} higher-order moments are given as
\begin{equation}
    \begin{split}
        \mathbb{E}[R^k] &= \left(\frac{\hat{r}}{\sqrt{2 \mu }}\right)^k (2 \mu )_{\frac{k}{2}}\\&\times \, _2F_1\left(\frac{2-k}{4},-\frac{k}{4};\mu
   +\frac{1}{2};\frac{H^2}{h^2}\right).
    \end{split}
\end{equation}
The mean and variance of the logarithm are obtained via numerical integration.
\subsection{Instantaneous and Mean SNR}

Let $R > 0$ denote the envelope of a fading signal. The instantaneous SNR at the receiver output is defined as
\begin{equation}
    \gamma(r) \triangleq \frac{E_b R^2}{N_0},
\end{equation}
where $E_b$ is the energy per bit and $N_0$ is the noise power spectral density. The average SNR is then given by
\begin{equation}
    \bar{\gamma} = \frac{E_b}{N_0} \mathbb{E}[R^2].
\end{equation}
Then,
\begin{equation}
    \gamma(r) = \bar{\gamma} \frac{R^2}{\mathbb{E}[R^2]}.
\end{equation}



 \subsection{The Product Distribution and the Lognormal}\label{subsec:prod_log}

There is an inherent connection between the distribution of the product of independent random variates and the Lognormal distribution. Let $Z = \prod_{i=1}^N X_i$, in which $\{X_i\}_{i=1}^{N}$ is a set of independent, positive random variates with finite log-variance. The logarithm of $Z$ can be written as the sum of the individual logarithms as $\log(Z) = \sum_{i=1}^N \log(X_i)$. Under these conditions, for large $N$, the central limit theorem holds and the distribution of $\log(Z)$ tends to the Normal distribution. Consequently, the distribution of $Z$ tends to the Lognormal in an exact manner with parameter $\nu = \mathbb{E}[\log(Z)]$ and $\sigma^2 = \mathbb{V}[\log(Z)]$.



The mapping process relating the product of random variables and a Lognormal variable may involve two distinct processes, depending on the target task, here identified as Reverse direction and Forward direction.\footnote{We borrow these denominations from the point-to-multipoint and multipoint-to-point communication systems.} In the Reverse direction, the aim is to map the product of RVs onto a Lognormal RV, which is a straightforward process. In Forward direction, the aim is to map the Lognormal RV onto the product of RVs, which constitutes a tricky and more interesting application. In the present work, we delve into these processes for the product of Nakagami-$m$ (Forward and Reverse), the I-Nakagami-$m$ (Forward and Reverse), and mixed products of arbitrary number of $\alpha$-$\mu$, $\kappa$-$\mu$, and $\eta$-$\mu$ variates (Reverse).

\section{The Product of Nakagami-$m$ and its Lognormal Mapping}
\label{sec: Nakagami and Lognormal Mapping}

This section models the Lognormal distribution using a product of Nakagami-$m$ variates. The corresponding logarithmic mean and variance are derived, and a moment-matching framework is proposed to relate the Nakagami-$m$ parameters to those of the target Lognormal distribution. In addition, two sample applications are presented based on this surrogate model. The first involves the derivation of asymptotically exact expressions for key performance metrics in wireless systems affected by Lognormal fading, including the CF, BER, and Shannon’s capacity. The second revisits the composite $\alpha$-$\mu$-Lognormal fading model to provide closed-form expressions for its PDF and CDF.

\subsection{From the Product of Nakagami-$m$ to Lognormal: 
The Reverse Direction}
\label{sec: product to Lognormal Nakagami}

Let $W_i$ be a \naka{} variate with mean power $\Omega_{W_i} = \mathbb{E}[W_i^2]$. Its PDF is defined as
\begin{equation}
    f_{W_i}(w) = \frac{2 m_i^{m_i}}{\Gamma(m_i)} \frac{w^{2m_i-1}}{{\Omega_{W_i}}^{m_i}}\exp\left(-\frac{m_i w^2}{\Omega_{W_i}}\right),
\end{equation}
in which $m_i > 0$ is related to the number of multipath clusters and is defined as the inverse of the normalized variance of $W_i$, i.e., $m_i = \mathbb{E}[W_i^2]^2 / \mathbb{V}[W_i^2]$. Now, let $X = \prod_{i=1}^N W_i$, then its PDF is given as \cite{Karagiannidis2007}
\begin{equation}\label{eq:prod_naka}
    f_X(x) {=} \frac{2}{x \Gamma_m} G_{0,N}^{N,0}\left[
    x^2  \frac{\prod_{i=1}^Nm_i}{\Omega_X}\left|
    \begin{smallmatrix}
    -\\
    m_1,...,m_N
    \end{smallmatrix}
    \right.\right],
\end{equation}
in which $\Omega_X = \mathbb{E}[X^2] = \prod_{i=1}^N\Omega_{W_i}$ is the mean power of $X$ and $\Gamma_m = \prod_{i=1}^N \Gamma(m_i)$. Its CDF is obtained in a straightforward manner as
\begin{equation}
    F_X(x) {=} \frac{1}{\Gamma_m} G_{1,N+1}^{N,1}\left[
    x^2  \frac{\prod_{i=1}^Nm_i}{\Omega_X}\left|
    \begin{smallmatrix}
    1\\
    m_1,...,m_N, 0
    \end{smallmatrix}
    \right.\right].
\end{equation}

The log-mean and log-variance are obtained, respectively, as \cite{Karagiannidis2007}
\begin{equation}\label{eq:naka_logmean}
    \nu_X  = \frac{1}{2} \left(\log \left(\frac{\Omega_X}{\prod_{i=1}^Nm_i} \right)+\sum _{i=1}^N \psi ^{(0)}\left(m_i\right)\right),
\end{equation}
and
\begin{equation}\label{eq:naka_logvar}
    \sigma_X^2 = \mathbb{V}[\log(X)]  = \frac{1}{4} \sum _{i=1}^N \psi ^{(1)}\left(m_i\right),
\end{equation}
The parameters $\nu_X$ and $\sigma_X$ are the equivalent parameters of the Lognormal that is reached from the product of $N$ Nakagami-$m$ variates.

\subsection{From the Lognormal to the Product of Nakagami-$m$: 
The Forward Direction
}\label{sec:map-log_prod}

As mentioned in Section \ref{subsec:prod_log}, the product of random variates tends in distribution to the Lognormal provided that the log-variance is finite which is the case for the product of \naka{}. In addition, finding a set of \naka{} variates that lead to a particular Lognormal distribution is of interest due to the PDF for the product being more analytically tractable than that of the Lognormal. Accomplishing this task requires finding the parameters, $m_i$ and $\Omega_{W_i}$, of $N$ \naka{} variates. In general, $2N$ relations between the product distribution and the \naka{} would be necessary, and as $N$ should be large solving this system could become infeasible. However, this complexity can be circumvented by arbitrarily imposing a fixed relation between the parameters of each \mbox{Nakagami-$m$} variate. The number of ways to parametrize this relationship is unbounded. Here, we will exercise the scenario in which we have an identical set of $m_i$. The variates are independent and non-identically distributed (i.n.i.d.) as the $\Omega_{W_i}$ might differ, but as can be seen in  \eqref{eq:prod_naka} all that is required is that $\Omega_X = \prod_{i=1}^N\Omega_{W_i}$, and without loss of generality, $\Omega_{W_i} = \Omega_X^{1/N}$. With these considerations, the number of unknowns is reduced to two, namely $m$ and $\Omega_X$. 

Again, there are several methods to obtain two relations between the product distribution and the Lognormal distribution. However, those that lead to the best approximation are the log-mean and log-variance.\footnote{Several methods can be used to relate the Lognormal distribution to its product-based approximation, including moment matching, moment-generating-function matching, and log-mean/log-variance matching (the approach adopted in this work). Although all asymptotically converge to the Lognormal law, the log-mean/log-variance method provides the most accurate and rapidly convergent method~\cite{Aitchison1957,Sapatnekar1992,Mehta2007}.
} 
In other words, we set the Lognormal parameters $\nu =\nu_X$ and $\sigma^2 = \sigma_X^2$, therefore
\begin{subequations}\label{eq:system_naka_iid}
\begin{align}
    &\nu =\frac{1}{2} (\log (\Omega_X )-N \log (m)+N \psi ^{(0)}(m)),\\
    &\sigma^2=\frac{N}{4} \psi ^{(1)}(m).
\end{align}
\end{subequations}
This system is easily solved numerically for $m$ and $\Omega_X$ after setting the desired number of variates, $N$, by any mathematical package such as Maple, MatLab, and Mathematica. Then the PDF of the Lognormal may be replaced with the more tractable one given in \eqref{eq:prod_naka}.

\subsection{Application I: Performance Analysis in Lognormal Fading}

In this subsection, the performance of a wireless system operating under Lognormal fading is evaluated. Specifically, using the product of Nakagami-$m$ random variables, asymptotically exact expressions for the CF, BER, and Shannon’s  capacity are derived.\footnote{In \cite{Yilmaz2009}, closed-form expressions for the BER, moment-generating function, and Shannon’s capacity were derived for the product of i.n.i.d. $\alpha$-$\mu$ variates---a generalization of the product of Nakagami-$m$ distributions. The corresponding expressions in our context can be obtained from their results through appropriate parameterization.}

\subsubsection{Characteristic Function}
The CF is the expected value of $\exp(j \omega r)$. For a Lognormal fading, this value has no closed-form, but an asymptotically exact, closed-form, expression may be obtained by using the PDF of the product of \naka{} with parameters obtained as described before. Now, after rewriting the $G$ function with its Mellin Barnes contour integral representation and using the Euler's formula, and changing the order of integration, the CF is obtained as
\begin{equation}
\begin{split}
    \phi_X(\omega) &= \frac{\sqrt{\pi}}{\Gamma(m)^N}\Bigg(G_{2,N}^{N,1}\left[\frac{4m^N}{\omega^2\Omega_X}\left|
    \begin{smallmatrix}
        1,1/2\\
        m,\ldots,m
    \end{smallmatrix}\right.\right]\\
    &+j \ \text{sign}(\omega) G_{2,N}^{N,1}\left[\frac{4m^N}{\omega^2\Omega_X}\left|
    \begin{smallmatrix}
        1/2,1\\
        m,\ldots,m
    \end{smallmatrix}\right.\right]\Bigg).
\end{split}
\end{equation}

\subsubsection{Bit Error Ratio}
The BER of a binary modulation scheme over a fading channel is obtained by averaging this metric with the channel's PDF. In \cite[Eq. (8.100)]{Simon:2005}, a general formula for the BER is given as $P(\gamma) = \Gamma(b, a \gamma(r)) / (2\Gamma(b))$\footnote{Refer to \cite[Table 8.1]{Simon:2005} for the appropriate values for $a$ and $b$.}. Then, in a Lognormal fading with parameters $\nu$ and $\sigma$, the BER is
\begin{equation}
    \bar{P}_b = \int_0^\infty P(\gamma) f_R(r)d r,
\end{equation}
Using \eqref{eq:pdf_lognormal}, this average is analytically intractable. Nonetheless, an asymptotically exact expression is obtained by using \eqref{eq:prod_naka} instead with parameters $m_i$ and $\Omega_{W_i}$ as described in Section \ref{sec:map-log_prod}. Now this integral is easily solved by using the Mellin-Barnes contour integral representation of the Meijer-G function, change the order of integration and reinterpreting the result as another Meijer-G function which results in
\begin{equation}
    \begin{split}
        \bar{P}_{b,X}=\frac{\Gamma (m)^{-N}}{2 \Gamma (b)} G_{2,N+1}^{N,2}\left[\frac{m^N}{a \bar{\gamma }}\left|
        \begin{smallmatrix}
            1,1-b\\
            m,\ldots,m, 0
        \end{smallmatrix}
        \right.\right].
    \end{split}
\end{equation}

\subsubsection{Shannon's Capacity}
The Shannon's capacity of a signal with bandwidth $B$ over AWGN channel is defined as $C(\gamma) = B \log_2(1+\gamma)$ in which $\gamma$ is the instantaneous SNR. Over fading channel, the average capacity is the expected value of $C(\gamma)$. In a Lognormal environment, no closed-form exists but an asymptotically exact expression can be obtained using the PDF for the product of \naka{} variates with parameters obtained as previously described. Therefore in a Lognormal fading the average capacity is
\begin{equation}
    \begin{split}
        \bar{C}_X = \frac{B}{\Gamma (m)^N \log (2)} G_{2,N+2}^{N+2,1}\left[\frac{m^N}{\bar{\gamma}}\left|\begin{smallmatrix}
            0,1\\
            0,0,m,\ldots,m
        \end{smallmatrix}\right.\right].
    \end{split}
\end{equation}


\subsection{Application II: The \am{}-Lognormal Fading Model}
\label{sec: a-u-Lognormal part1}

Here, we revisit the composite $\alpha$-$\mu$-Lognormal fading model. 
This model captures two fundamental propagation effects that are ubiquitous in wireless communication systems: multipath fading and shadowing. In this formulation, the multipath component is modeled using the $\alpha$-$\mu$ distribution, while the shadowing effect is represented by a Lognormal distribution that governs the fluctuations in the mean power.
Accordingly, the composite (multipath-shadowing) distribution is obtained by averaging the conditional PDF of the $\alpha$-$\mu$ distribution over the Lognormal distribution of the mean power. Let $R$ denote the envelope of the $\alpha$-$\mu$ fading component such that $\mathbb{E}[R^2] = \delta \, \Omega_R$, where $\delta$ is a lognormally distributed random variable representing the shadowing component. Then, the conditional PDF of $R$ given $\delta$ can be written as
\begin{equation}\label{eq:au_delta}
    \begin{split}
        f_{R|\delta}(r|\delta){=}\frac{\alpha  r^{\alpha  \mu -1}}{\delta ^{\frac{\alpha  \mu
   }{2}} \Gamma (\mu )} \left({\frac{(\mu )_{{2}/{\alpha
   }}}{\Omega_R }}\right)^{\frac{\alpha  \mu}{2} } e^{-\frac{r^{\alpha }}{\delta
   ^{\alpha /2}} \left({\frac{(\mu )_{{2}/{\alpha }}}{\Omega_R
   }}\right)^{\frac{\alpha}{2} }}.
    \end{split}
\end{equation}

If one attempts to integrate \eqref{eq:au_delta} over $\delta$ assuming a Lognormal PDF, the resulting expression does not admit a closed-form solution and may become analytically intractable. To overcome this limitation, as discussed in Sections~\ref{subsec:prod_log} and \ref{sec: product to Lognormal Nakagami}, we approximate the Lognormal shadowing component by modeling it as the product of a Nakagami-$m$ random variable with parameters $m$ and $\Omega_X$. This surrogate model enables analytical tractability while preserving the essential statistical behavior of the original composite model. Under this approximation, the integral over $\delta$ can be evaluated in closed form using the Fox-H function \cite[Eq. (1.2)]{Mathai:2009}. Consequently, an asymptotically exact expression for the PDF of the composite $\alpha$-$\mu$-Lognormal fading model is obtained as
\begin{equation}
    \label{eq: PDF a-u-Lognormal}
    \begin{split}
        f_{R,X}(r) &= \frac{4 }{r \Gamma (\mu )\Gamma (m)^{N}} \\
        &\!\!\!\!\!\!\!\!\!\!\!\!\!\!\!\!\!\times H_{0,N+1}^{N+1,0}\left[\frac{r^4 m^N (\left(\mu\right)_{{2}/{\alpha }})^2
   }{\left(\Omega _R\right)^2 \Omega_X }
        \left|
        \begin{smallmatrix}
            -\\
            (m,1),\ldots,(m,1), (\mu,\frac{4}{\alpha})
        \end{smallmatrix}
        \right.\right]\!.
    \end{split}
\end{equation}

On the other hand, an asymptotically exact
expression for the CDF can be obtained by integrating \eqref{eq: PDF a-u-Lognormal} from zero to $r$, yielding
\begin{equation}
    \label{eq: CDF a-u-Lognormal}
    \begin{split}
        F_{R,X}(r) &= \frac{1}{\Gamma (\mu )\Gamma (m)^{N}} \\
        &\!\!\!\!\!\!\!\!\!\!\!\!\!\!\!\!\!\!\!\!\!\!\times H_{1,N+2}^{N+1,1}\left[\frac{r^4 m^N (\left(\mu\right)_{\frac{2}{\alpha }})^2
   }{\left(\Omega _R\right)^2 \Omega_X }
        \left|
        \begin{smallmatrix}
            (1,1)\\
            (m,1),\ldots,(m,1), (\mu,\frac{4}{\alpha}),(0,1)
        \end{smallmatrix}
        \right.\right]\!.
    \end{split}
\end{equation}
Observe that when $\alpha = 4$, the Fox-H function in both the PDF and CDF expressions simplifies to a Meijer-G Function, thus the PDF and CDF for this environment are obtained, respectively, as
\begin{equation}
    \begin{split}
        f_{R,X}(r) &= \frac{4 }{r \Gamma (\mu )\Gamma (m)^{N}} \\
        &\times G_{0,N+1}^{N+1,0}\left[\frac{r^4 m^N (\left(\mu\right)_{\frac{1}{2 }})^2
   }{\left(\Omega _R\right)^2 \Omega_X }
        \left|
        \begin{smallmatrix}
            -\\
            m,\ldots,m, \mu
        \end{smallmatrix}
        \right.\right]\!
    \end{split}
\end{equation}
\begin{equation}
    \begin{split}
        F_{R,X}(r) &= \frac{1}{\Gamma (\mu )\Gamma (m)^{N}} \\
        &\times G_{1,N+2}^{N+1,1}\left[\frac{r^4 m^N (\left(\mu\right)_{\frac{1}{2}})^2
   }{\left(\Omega _R\right)^2 \Omega_X }
        \left|
        \begin{smallmatrix}
            1\\
            m,\ldots,m, \mu,0
        \end{smallmatrix}
        \right.\right]\!.
    \end{split}
\end{equation}

As for the case when $\alpha = 2$, a simplification is found by applying the Gamma's duplication formula \cite[Eq. (6.1.18)]{Abramowitz:1972}, which results in the following for the PDF and CDF:
\begin{equation}
     \begin{split}
        f_{R,X}(r) &= \frac{2^{\mu +1}}{r \sqrt{\pi } \Gamma (\mu ) \Gamma (m)^N} \\
        &\times G_{0,N+2}^{N+2,0}\left[\frac{r^4 m^N \mu ^2}{4 \Omega _R^2 \Omega _X}
        \left|
        \begin{smallmatrix}
            -\\
            m,\ldots,m, \frac{\mu}{2},\frac{\mu+1}{2}
        \end{smallmatrix}
        \right.\right]\!
    \end{split}
\end{equation}
\begin{equation}
     \begin{split}
        F_{R,X}(r) &= \frac{2^{\mu -1}}{\sqrt{\pi } \Gamma (\mu ) \Gamma (m)^N} \\
        &\times G_{1,N+3}^{N+2,1}\left[\frac{r^4 m^N \mu ^2}{4 \Omega _R^2 \Omega _X}
        \left|
        \begin{smallmatrix}
            1\\
            m,\ldots,m, \frac{\mu}{2},\frac{\mu+1}{2},0
        \end{smallmatrix}
        \right.\right]\!.
    \end{split}
\end{equation}

It is worth highlighting that the $\alpha$-$\mu$-Lognormal distribution encompasses several Lognormal-based composite fading models as special cases, including the Nakagami-Lognormal \cite{Hussaini02}, Weibull-Lognormal \cite{Karadimas09}, Rayleigh-Lognormal (Suzuki distribution) \cite{Hansen77}.
Notably, to date, no exact closed-form expressions are available for the main statistical functions (namely, the PDF and CDF) of these distributions. 
In this context, the expressions derived in \eqref{eq: PDF a-u-Lognormal} and \eqref{eq: CDF a-u-Lognormal}, along with those presented in \eqref{eq: PDF a-u-Lognormal I-Nakagami} and \eqref{eq: CDF a-u-Lognormal I-Nakagami} in subsequent sections, establish a unified and effective analytical framework for characterizing the statistical behavior of Lognormal-based composite fading models. 

\section{The Product of I-Nakagami-$m$ and its Lognormal Mapping}
\label{sec: I-Nakagami and Lognormal Mapping}

This section introduces an alternative surrogate model for the Lognormal distribution, constructed from the product of I-Nakagami-$m$ variates. A moment-matching approach is employed to relate the parameters of the surrogate model to those of the target Lognormal distribution. Additionally, two sample applications are presented based on this model. The first involves the derivation of asymptotically exact expressions for key performance metrics in wireless systems subject to Lognormal fading. The second capitalizes on the previous introduction of the \am{}-Lognormal and provides asymptotically exact expressions for the PDF and CDF as a function of the product of I-Nakagami-$m$.




\subsection{From the Product of I-Nakagami-$m$ to the Lognormal: The Reverse Direction}
Let $T_i = 1 / W_i$ such that $W_i$ is a Nakagami-\emph{m} variate as described before, with $m_i>1$, and let $\Omega_{T_i} = \mathbb{E}[T_i^2] = m_i / (\Omega_{W_i}(m_i-1))$, then the PDF of the I-Nakagami-$m$ is easily obtained as
\begin{equation}
    f_{T_i}(t)=\frac{2 \left(\left(m_i-1\right) \Omega _{T_i}\right){}^{m_i}}{\Gamma \left(m_i\right) t^{2 m_i+1}} e^{-\frac{\left(m_i-1\right)
   \Omega _{T_i}}{t^2}}.
\end{equation}
Now, let $Y=\prod_{i=1}^NT_i$ be the product of $N$ independent arbitrarily distributed I-Nakagami-$m$ variates. By following a similar approach as in \cite{Karagiannidis2007}, the PDF and CDF of $Y$ is obtained, respectively, as 
\begin{equation}\label{eq:prod_inv}
\begin{split}
    f_Y(y)&=\frac{2}{\sqrt{\Omega _Y}\Gamma_m \prod _{i=1}^N  \sqrt{m_i-1}}\\
    &\!\!\!\!\!\!\times G_{N,0}^{0,N}\left[\frac{y^2}{\Omega _Y \prod _{i=1}^N \left(m_i-1\right)} \left|\begin{smallmatrix}
        \frac{1}{2}-m_1,\ldots , \frac{1}{2}-m_N\\-
    \end{smallmatrix}\right.
    \right]
\end{split}
\end{equation}
\begin{equation}
    \begin{split}
        &F_Y(y)=\frac{y}{\sqrt{\Omega _Y} \Gamma_m\prod _{i=1}^N \sqrt{m_i-1}}\\
    &\times G_{N+1,1}^{0,N+1}\left[\frac{y^2}{\Omega _Y \prod _{i=1}^N \left(m_i-1\right)} \left|\begin{smallmatrix}
        \frac{1}{2}-m_1,\ldots , \frac{1}{2}-m_N, \frac{1}{2}\\-\frac{1}{2}
    \end{smallmatrix}\right.
    \right],
    \end{split}
\end{equation}
in which $\Omega_Y = \mathbb{E}[Y^2] = \prod_{i=1}^N \Omega_{T_i}$ is the mean power of the random variate $Y$ and $\Gamma_m$ is defined previously in Section \ref{sec: product to Lognormal Nakagami}. For sufficiently large $N$, this PDF will tend in distribution to the Lognormal with parameters obtained from the logarithmic mean of $Y$ and the logarithmic variance of $Y$, which are obtained as
\begin{equation}
    \nu _Y=\frac{\log \left({\Omega _Y}\right)}{2}+\sum _{i=1}^N \left(\frac{\log \left({m_i-1}\right)}{2}-\frac{\psi ^{(0)}\left(m_i\right)}{2}\right),
\end{equation}
and
\begin{equation}
    \sigma _Y^2=\frac{1}{4} \sum _{i=1}^N \psi ^{(1)}\left(m_i\right).
\end{equation}
%
%
%
\subsection{From the Lognormal to the Product of I-Nakagami-\emph{m}: The Forward Direction}\label{sec:log-to-inv}
Similarly to Subsection~\ref{sec:map-log_prod}, consider an arbitrary Lognormal distribution with parameters $\nu$ and $\sigma$. The goal is to determine a set of I-Nakagami-$m$ variates that converges to this distribution. Specifically, this involves identifying $N$ shape parameters $m_i$ and $N$ scale parameters $\Omega_{T_i}$ such that their product approximates the target Lognormal behavior. However, for large $N$, this process may become computationally impractical for arbitrary parameter values.
Therefore, we restrict this analysis for the independent and identically distributed (i.i.d.) scenario, i.e., $m_i = m$ and $\Omega_{T_i} = \Omega_Y^{1/N}$. Now, two relations between the Lognormal distribution and the product distribution are required. Let's consider the logarithmic mean and the logarithmic variance, then we set $\nu = \nu_Y$ and $\sigma^2 = \sigma_Y^2$, 
\begin{subequations}\label{eq:system_I_naka_iid}
\begin{align}
    &\nu = \frac{N}{2} \left(\log \left((m-1) \Omega _Y^{1/N}\right)-\psi ^{(0)}(m)\right),\\
    &\sigma^2 =\frac{N \psi ^{(1)}(m)}{4}.
\end{align}
\end{subequations}
This system of equations is easily solved numerically for $m$ and $\Omega_Y$ after choosing the desired number of variates.


\subsection{Application I: Performance Analysis in Lognormal Fading}

In this subsection, the performance of a wireless system subject to Lognormal fading is evaluated using the surrogate model based on the product of I-Nakagami-$m$ random variables. Asymptotically closed-form expressions for the CF, BER, and Shannon's capacity are derived.

\subsubsection{Characteristic Function}
The CF for the Lognormal fading can be asymptotically reached in closed-form by averaging $\exp(j \omega r)$ with the PDF the product of I-Nakagami-$m$ using the parameter estimation as described in Subsection \ref{sec:log-to-inv}. Then, the CF is obtained as in \eqref{eq:cf_inv} at the top of the next page.
\begin{figure*}[ht!]
\begin{equation}\label{eq:cf_inv}
    \phi_Y(\omega) {=} \frac{2 \sqrt{\pi } \Gamma (m)^{-N}}{\sqrt{(m{-}1)^N \Omega _Y}}\Bigg(\frac{1}{\left|\omega\right|}G_{N+2,0}^{0,N+1}\left[\frac{4 (m{-}1)^{-N}}{\Omega _Y \omega ^2}\left|\begin{smallmatrix}
        \frac{1}{2}, \frac{1}{2}-m,\ldots,\frac{1}{2}-m,0\\
        -
    \end{smallmatrix}\right.\right] {+} \frac{j}{\omega}G_{N+2,0}^{0,N+1}\left[\frac{4 (m{-}1)^{-N}}{\Omega _Y \omega ^2}\left|\begin{smallmatrix}
    0, \frac{1}{2}-m,\ldots,\frac{1}{2}-m,\frac{1}{2}\\
        -
    \end{smallmatrix}\right.\right]\Bigg)
\end{equation}
\hrulefill
\end{figure*}

\subsubsection{Bit Error Rate}
A closed, asymptotically exact, formulation for the BER for a binary modulation in a Lognormal channel approximated by the product of I-Nakagami-$m$ variates can be obtained by averaging the BER in a Gaussian channel with \eqref{eq:prod_inv}. The resulting integral is solved by first writing the Meijer-G function with its Mellin-Barnes contour integration and then change the order of integration, which results in
\begin{equation}
    \begin{split}
        \bar{P}_{b,Y}&=\frac{\Gamma (m)^{-N}}{2 \Gamma (b) \sqrt{a (m-1)^N \bar{\gamma }}}\\
        &\times G_{N+2,1}^{0,N+2}\left[\frac{(m-1)^{-N}}{a \bar{\gamma }}\left|\begin{smallmatrix}
            \frac{1}{2}-b,\frac{1}{2},\frac{1}{2}-m,\ldots,\frac{1}{2}-m\\
            -\frac{1}{2}
        \end{smallmatrix}\right.\right].
    \end{split}
\end{equation}

\subsubsection{Shannon's Capacity}
The Shannon's Capacity in a fading environment is the average of $C(\gamma) = B log_2( 1+\gamma(r))$. In a Lognormal scenario, exact analytical expressions are hard, if not unfeasible, to obtain. However, by replacing the Lognormal PDF with the PDF for the product of I-Nakagami-$m$, an asymptotically exact expression for the capacity is obtained as
\begin{equation}
\begin{split}
    \bar{C}_Y&=\frac{B\ \Gamma (m)^{-N}}{\log (2) \sqrt{(m-1)^N \bar{\gamma }}} \\
    &\times G_{N+2,2}^{2,N+1}\left[\frac{(m-1)^{-N}}{\bar{\gamma }}\left|
    \begin{smallmatrix}
        \frac{1}{2}-m,\ldots,\frac{1}{2}-m,-\frac{1}{2},\frac{1}{2}\\
        -\frac{1}{2},-\frac{1}{2}
    \end{smallmatrix}
    \right.\right].
\end{split}
\end{equation}
%
\subsection{Application II: The \am{}-Lognormal Fading Model}

In this section, we derive asymptotically exact expressions for the PDF and CDF of the $\alpha$-$\mu$-Lognormal fading model.
While the overall derivation strategy follows the methodology previously outlined, a key distinction lies in the alternative approximation adopted for the Lognormal shadowing component. Specifically, the Lognormal distribution is approximated using the product of independent I-Nakagami-$m$ random variables. By leveraging this surrogate model---together with the Fox-H function framework introduced in Section~\ref{sec: a-u-Lognormal part1}---we obtain closed-form, asymptotically exact expressions for the PDF and CDF, given respectively by
\begin{equation}
    \label{eq: PDF a-u-Lognormal I-Nakagami}
    \begin{split}
        f_{R,Y}(r)&=\frac{4 r \Gamma (m)^{-N} (\mu )_{\frac{2}{\alpha }}}{\Gamma (\mu )
   \Omega _R \sqrt{(m-1)^N \Omega _Y}}\\
   &\!\!\!\!\!\!\!\!\!\!\!\!\!\!\times H_{N,1}^{1,N}\left[\frac{r^4 \left((\mu )_{\frac{2}{\alpha }}\right)^2}{\Omega _R^2 \Omega _Y
   (m-1)^N}\left|\begin{smallmatrix}
       \left(\frac{1}{2}-m,1\right),\ldots,\left(\frac{1}{2}-m,1\right)\\
       \left(\mu-\frac{2}{\alpha},\frac{4}{\alpha}\right)
   \end{smallmatrix}\right.\right]
    \end{split}
\end{equation}
\begin{equation}
    \label{eq: CDF a-u-Lognormal I-Nakagami}
    \begin{split}
        F_{R,Y}(r)&=\frac{r^2 \Gamma (m)^{-N} (\mu )_{\frac{2}{\alpha }}}{\Gamma (\mu )
   \Omega _R \sqrt{(m-1)^N \Omega _Y}}\\
   &\!\!\!\!\!\!\!\!\!\!\!\!\!\!\!\!\!\!\!\!\times H_{N+1,2}^{1,N+1}\left[\frac{r^4 \left((\mu )_{\frac{2}{\alpha }}\right)^2}{\Omega _R^2 \Omega _Y
   (m-1)^N}\left|\begin{smallmatrix}
       \left(\frac{1}{2}-m,1\right),\ldots,\left(\frac{1}{2}-m,1\right),(\frac{1}{2},1)\\
       \left(\mu-\frac{2}{\alpha},\frac{4}{\alpha}\right),(-\frac{1}{2},1)
   \end{smallmatrix}\right.\right]\!.
    \end{split}
\end{equation}
For the special case $\alpha = 4$, these expressions simplify and can be written in terms of the more tractable Meijer-G function, as
\begin{equation}
    \begin{split}
        f_{R,Y}(r)&=\frac{4 r \Gamma (m)^{-N} (\mu )_{\frac{1}{2}}}{\Gamma (\mu )
   \Omega _R \sqrt{(m-1)^N \Omega _Y}}\\
   &\times G_{N,1}^{1,N}\left[\frac{r^4 \left((\mu )_{\frac{1}{2 }}\right)^2}{\Omega _R^2 \Omega _Y
   (m-1)^N}\left|\begin{smallmatrix}
       \frac{1}{2}-m,,\ldots,\frac{1}{2}-m\\
       \mu-\frac{2}{\alpha}
   \end{smallmatrix}\right.\right]
    \end{split}
\end{equation}
\begin{equation}
    \begin{split}
        F_{R,Y}(r)&=\frac{r^2 \Gamma (m)^{-N} (\mu )_{\frac{1}{2 }}}{\Gamma (\mu )
   \Omega _R \sqrt{(m-1)^N \Omega _Y}}\\
   &\!\!\!\!\!\!\!\!\!\times G_{N+1,2}^{1,N+1}\left[\frac{r^4 \left((\mu )_{\frac{1}{2 }}\right)^2}{\Omega _R^2 \Omega _Y
   (m-1)^N}\left|\begin{smallmatrix}
       \frac{1}{2}-m,\ldots,\frac{1}{2}-m,\frac{1}{2}\\
       \mu-\frac{2}{\alpha},-\frac{1}{2}
   \end{smallmatrix}\right.\right]\!.
    \end{split}
\end{equation}

For $\alpha = 2$, the Gamma's duplication formula can be applied to simplify the expressions for the PDF and CDF, yielding,
\begin{equation}
    \begin{split}
        f_{R,Y}(r)&=\frac{2^{\mu } \Gamma (m)^{-N} \mu\  r}{\sqrt{\pi } \Gamma (\mu ) \Omega _R \sqrt{(m-1)^N \Omega _Y}}\\
   &\!\!\!\!\!\!\!\!\!\!\!\!\!\!\times G_{N,2}^{2,N}\left[\frac{r^4 \mu^2}{4\Omega _R^2 \Omega _Y
   (m-1)^N}\left|\begin{smallmatrix}
       \frac{1}{2}-m,\ldots,\frac{1}{2}-m\\
       \frac{\mu-1}{2},\frac{\mu}{2}
   \end{smallmatrix}\right.\right]
    \end{split}
\end{equation}
\begin{equation}
    \begin{split}
        F_{R,Y}(r)&=\frac{2^{\mu -2} \Gamma (m)^{-N} \mu\  r^2}{\sqrt{\pi } \Gamma (\mu ) \Omega _R \sqrt{(m-1)^N \Omega _Y}}\\
   &\!\!\!\!\!\!\!\!\!\!\!\!\!\!\times G_{N+1,3}^{2,N+1}\left[\frac{r^4 \mu^2}{4\Omega _R^2 \Omega _Y
   (m-1)^N}\left|\begin{smallmatrix}
       \frac{1}{2}-m,\ldots,\frac{1}{2}-m,\frac{1}{2}\\
       \frac{\mu-1}{2},\frac{\mu}{2},-\frac{1}{2}
   \end{smallmatrix}\right.\right]\!.
    \end{split}
\end{equation}

\section{Convergence Improvement}
\label{sec: Convergence Improvement}

Although the surrogate models presented in Sections~\ref{sec: Nakagami and Lognormal Mapping} and~\ref{sec: I-Nakagami and Lognormal Mapping} yield asymptotically exact expressions, their accuracy depends on the number of product terms used. 
In practical scenarios, computational constraints may limit the feasible number of product terms, potentially degrading the approximation accuracy. To overcome this limitation, we utilize a random mixture strategy that exploits the complementary behavior of the Nakagami-$m$ and I-Nakagami-$m$ product distributions, enabling the same level of accuracy with markedly fewer multiplicative variates. Although convergence in distribution is theoretically guaranteed, the proposed hybrid formulation substantially improves the convergence rate in practice, thereby enabling a successful mapping with markedly fewer multiplicative terms and enhanced approximation fidelity.

The Nakagami-\emph{m} and the I-Nakagami-\emph{m} distributions exhibit a noteworthy complementary behavior when expanded into series representations, such as Taylor series. Specifically, the Nakagami-\emph{m} distribution offers faster convergence for low envelope values, while the I-Nakagami-\emph{m} counterpart converges more effectively for high envelope values. This duality is inherited by their product distributions. Consequently, series expansions based on Nakagami-\emph{m} variates are more accurate in low-envelope regimes, whereas those using I-Nakagami-\emph{m} are more efficient in high-envelope scenarios. This observation motivates the conjecture that a judicious combination of both---via a random mixture involving the Nakagami-\emph{m} and I-Nakagami-\emph{m} distributions---could significantly reduce the number of terms required to approximate a target Lognormal distribution with high accuracy. This reduction in complexity makes the approach especially attractive for applications where computational efficiency is critical.

Let $Z = X^q Y^{1-q}$, where $X$ is the product of $N$ Nakagami-$m$ variates and $Y$ is the product of $N$ I-Nakagami-\emph{m} variates, and $q$ is a Bernoulli variable that takes the value $1$ with probability $p$ and $0$ with probability $1-p$. Thus, each realization of $Z$ corresponds either to the product of Nakagami-\emph{m}  variates (with probability $p$) or to the product of I-Nakagami-\emph{m} variates (with probability $1-p$). From standard probability theory, the PDF of $Z$ is expressed as a combination of the PDFs of $X$ and $Y$, namely,
\begin{equation}
    f_Z(z) = p f_X(z) + (1-p) f_Y(z).
\end{equation}
In fact, any linear transformation of the PDF of $Z$---including the CF, BER, and composite fading distributions---can be obtained from the equivalent transformation of the PDF of $X$ and $Y$. Let $g_Z$, $g_X$ and $g_Y$ denote such transformations applied to $Z$, $X$ and $Y$, respectively. Then, it follows that
\begin{equation}
    g_Z = p\ g_X + (1-p)g_Y.
\end{equation}

As will be demonstrated in Section~\ref{sec: Some Plots}, the random mixture approach introduced in this section further enhances statistical fidelity while requiring markedly fewer multiplicative variates. 

\section{Some Further Useful Lognormal Mappings}\label{sec:usefull}


As already mentioned, the product of independent positive variates with finite log-variances tends to a Lognormal distribution as the number of variates tends to infinity. Hence, the approach proposed in \cite{10883648}, and enhanced here can be applied to many other scenarios of interest. For instance, products of random variates are found in several wireless applications, such as multihop links, RIS assisted links. The fading model in each link of a cascaded channel might differ. Let's consider the following resultant fading signal
\begin{equation}
    R = \underbrace{\prod_{i=1}^N X_i}_{X}\ \underbrace{\prod_{j=1}^MY_i}_{Y}\ \underbrace{\prod_{n=1}^LZ_i}_{Z},
\end{equation}
and $X_i$, $Y_i$, and $Z_i$ are \am{}, \km{} and \et{} distributed variates respectively. If the number of variates grows then the PDF of $R$ tends to the Lognormal distribution with parameters $\nu = \mathbb{E}[\log(R)]$ and $\sigma^2=\mathbb{V}[\log(R)]$. It is not difficult to show that $\nu = \mathbb{E}[\log(X)]+\mathbb{E}[\log(Y)]+\mathbb{E}[\log(Z)]$ and $\sigma^2 = \mathbb{V}[\log(X)]+\mathbb{V}[\log(Y)]+\mathbb{V}[\log(Z)]$. In other words, the Lognormal parameters could be obtained individually for each parcel in the product, by first approximating the product of \am{}, followed by the product of \km{} and then the product of \et{}.

In this section, we shall illustrate how to map the product of \am{}, \km{} and \et{} variates individually and their mixed combinations onto their corresponding Lognormal distributions (Reverse direction). The Forward direction, i.e. from Lognormal to their individual or combined products is of no interest since their corresponding PDFs are given in infinite series forms. On the other hand, once the Lognormal parameters are given in terms of the $\alpha$, $\eta$, $\kappa$ parameters, the statistics of interest can be obtained as in Section \ref{sec:map-log_prod} and Section \ref{sec:log-to-inv}. Additionally, as exercised previously for the Nakagami-$m$ and I-Nakagami-$m$, convergence can be improved if direct and inverse PDFs are combined. 

\subsection{Mapping the Product of \am{} Random Variables to the Lognormal - the Reverse Direction}
Let $X = \prod_{i=1}^NX_i$ such that $X_i$ are \am{} random variables with parameters \{$\alpha_i,\mu_i,\hat{r}_i$\}. As the number of variates enlarges, the PDF of $X$ will tend to the Lognormal. The Lognormal parameters may be obtained as
\begin{equation}
    \nu_X = \sum _{i=1}^N \left(\log \left(\hat{r}_i\right)+\frac{\psi ^{(0)}\left(\mu _i\right)-\log \left(\mu _i\right)}{\alpha _i}\right),
\end{equation}
and 
\begin{equation}
    \sigma_X^2 = \sum _{i=1}^N \frac{\psi ^{(1)}\left(\mu _i\right)}{\alpha _i^2}.
\end{equation}

\subsection{Mapping the Product of \km{} Random Variables to the Lognormal - the Reverse Direction}
Let $Y = \prod_{i=1}^M Y_i$ such that $Y_i$ are \km{} random variables with parameters \{$\kappa_i, \mu_i, \hat{r}_i$\}. The Lognormal that results from this product will have parameters given as
\begin{equation}
\begin{split}
    \nu_Y=\sum _{i=1}^M \nu_\kappa^{(i)},
\end{split}
\end{equation}
and 
\begin{equation}
    \begin{split}
        &\sigma_Y^2 = \sum _{i=1}^M \left(\mathbb{E}\left[\log ^2\left(y_i\right)\right]-\left(\nu_\kappa^{(i)}\right)^2\right),
    \end{split}
\end{equation}
in which $\nu_\kappa^{(i)}$ is defined in \eqref{eq:ku_logmean}. Alternatively, to circumvent the use of numerical integration, a moment matching approach may be employed. Let $K$ be the $k$-th moment of $Y$, i.e., $K = \prod_{i=1}^M\mathbb{E}[Y_i^k]$ in which $\mathbb{E}[Y_i^k]$ is as provided in \eqref{eq:ku_moment}, then the other parameter may be obtained as
\begin{equation}
    \sigma_Y^2=\frac{2 \log (K)-2 k\ \nu _Y}{k^2},
\end{equation}
after matching $K$ with \eqref{eq:log_moment}.
The higher-order moment of the product may rise quickly as the order increases. Thus it might be interesting to use values of $k$ that are closer to zero. As a useful and functional rule of thumb, the moment $k = 1/M$ can be applied.

\subsection{Mapping the Product of \et{} Random Variable to the Lognormal - the Reverse Direction}
Let $Z = \prod_{i=1}^LZ_i$ such that $Z_i$ are \et{} random variables with parameters \{$\eta_i, \mu_i,\hat{r}_i$\}. In this case both the logarithmic mean and logarithmic variance are not found in closed form but they can be easily obtained through numerical integration. Again, moment matching can be used to obtain both parameters in closed form. Here two moments are required. Let $K_1 = \prod_{i=1}^L\mathbb{E}[Z_i^{k_1}]$ and $K_2= \prod_{i=1}^L\mathbb{E}[Z_i^{k_2}]$ the $k_1$-th and $k_2$-th moment of the $Z$. After matching the moments of the product with those of the Lognormal given in \eqref{eq:log_moment}, the parameters are obtained as
\begin{equation}
    \nu_\eta = \frac{k_1^2 \log \left(K_2\right)-k_2^2 \log \left(K_1\right)}{k_1 k_2 \left(k_1-k_2\right)},
\end{equation}
and
\begin{equation}
\sigma_Z^2=\frac{2 k_1 \log \left(K_2\right)-2 k_2 \log \left(K_1\right)}{k_1 k_2 \left(k_2-k_1\right)}.   
\end{equation}
Again, it might be more appropriate to choose $k_1$ and $k_2$ close to zero due to the rapid growth of the moments of the product. As a rule of thumb, the moments $k_1 = 1/L$ and $k_2 = 2/L$ are useful and functional.
 
\subsection{Possible RIS Application}
In an RIS system, the typical channel model is defined as the sum of the products of the channel gains from the transmitter to each reflective element and from each reflective element to the receiver. As the signal propagates through the environment, the dynamic nature of the wireless medium may alter its statistical characterization, and each component of the model corresponds to the product of random variates. In other words, the overall scenario traversed by the signal may comprise different environments, each of which is described by distinct statistics, with the overall statistics given by the product of the individual ones. As the number of terms increases, this product can be approximated by a Lognormal distribution, as suggested in Section~\ref{sec:usefull}. The performance metrics in this case boil down to the sum of Lognormal variates, which remains as an open issue, and for which no closed-form solution is available to date. However, several approximation techniques have been proposed to characterize the PDF and CDF of the sum of Lognormals, the most conventional being to approximate the sum of Lognormal random variates by another Lognormal variate with parameters obtained through the method of Schwartz and Yeh~\cite{Yeh82}. A detailed analysis of the sum of Lognormal variates, however, lies beyond the scope of this work.

\section{Some Plots} \label{sec: Some Plots}

In this section, a few plots are presented to corroborate our findings. In all figures, two representative sets of Lognormal variables are used to illustrate a statistical metric. The first set has parameters $\nu = \sigma = 0.5$ while the second has $\nu = -1$ and $\sigma=1$. 

\begin{figure}[!t]
    \centering
    \includegraphics[width=\linewidth]{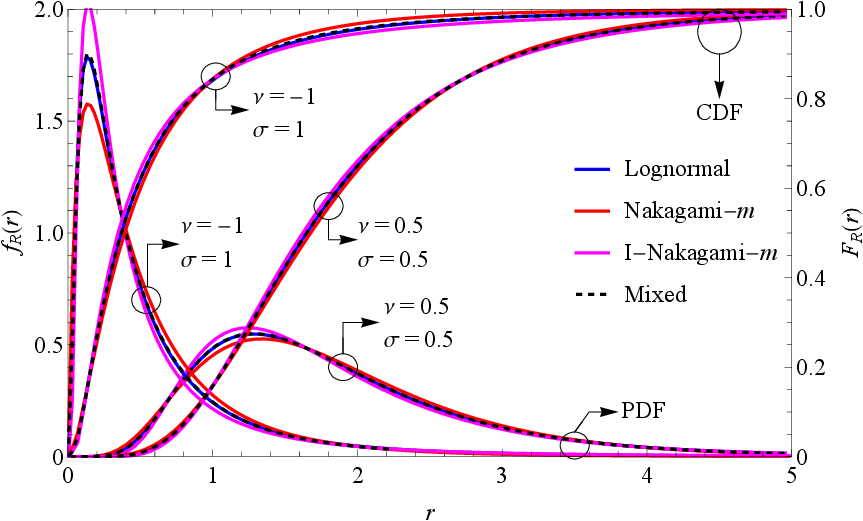}
    \caption{Lognormal PDF and its Nakagami-\emph{m}, I-Nakagami-\emph{m} and the mixed distribution.}
    \label{fig:PDFCDF}
\end{figure}

In Figure \ref{fig:PDFCDF} the PDF and CDF of the Lognormal are compared with that of the product of five Nakagami-$m$, five I-Nakagami-$m$ ($N = 5$), a relatively small number of variates,  whose parameters were obtained by following the procedures described in Section \ref{sec:map-log_prod} and \ref{sec:log-to-inv}, respectively, and the mixture model with $p = 0.5$. The Nakagami-$m$ and I-Nakagami-$m$ parameters obtained are shown in Table \ref{tab:logtoprod}. As it can be seen, 
for a small number of variates, the individual products of Nakagami-$m$ or I-Nakagami-$m$ fail to adhere to the exact Lognormal curve. Although, the mixed model matches the Lognormal exactly. 

\begin{figure}[!t]
    \centering
    \includegraphics[width=\linewidth]{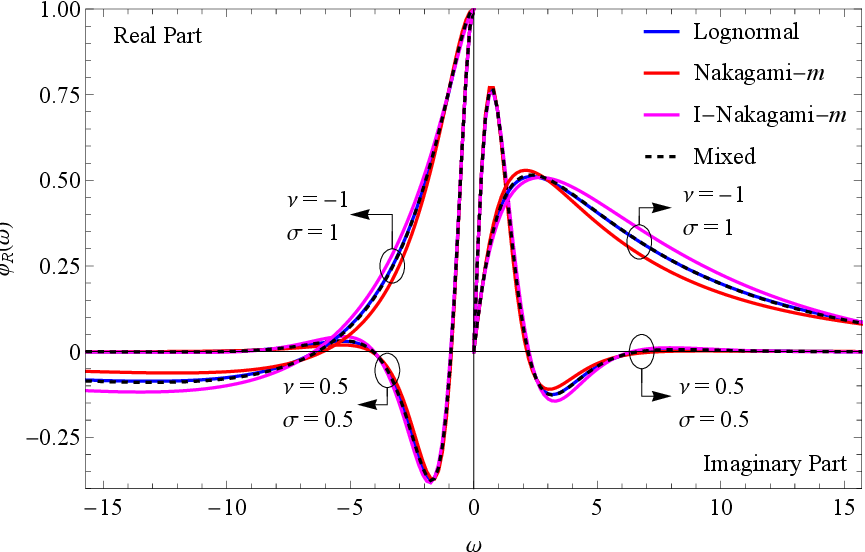}
    \caption{Lognormal PDF and its Nakagami-\emph{m}, I-Nakagami-\emph{m} and the mixed distribution.}
    \label{fig:CF}
\end{figure}

In Figure \ref{fig:CF}, the real and imaginary part of the Lognormal CF are depicted with the Nakagami-\emph{m} and the I-Nakagami-\emph{m}, again the mixture model shows great adherence to exact CF of the Lognormal. Note that the positive frequencies depict the CF's imaginary part, while the negative frequencies depict the CF's real part. This representation is sufficient, as the real part is an even function and the imaginary part is an odd function.

\begin{table}[ht]
\centering
\caption{Parameters for the Nakagami-$m$ and I-Nakagami-$m$}
\label{tab:logtoprod}
\resizebox{\columnwidth}{!}{%
\begin{tabular}{@{}ccccccccc@{}}
\toprule
\multirow{3}{*}{Lognormal} & \multicolumn{4}{c}{Nakagami-$m$}                       & \multicolumn{4}{c}{I-Nakagami-$m$}                     \\ \cmidrule(l){2-9} 
                           & \multicolumn{2}{c}{$N=5$} & \multicolumn{2}{c}{$N=10$} & \multicolumn{2}{c}{$N=5$} & \multicolumn{2}{c}{$N=10$} \\ \cmidrule(l){2-9} 
                           & $m$      & $\Omega_X$     & $m$      & $\Omega_X$      & $m$      & $\Omega_Y$     & $m$      & $\Omega_Y$      \\ \midrule
$\nu = 0.5$ &
  \multirow{2}{*}{5.48} &
  \multirow{2}{*}{4.35} &
  \multirow{2}{*}{10.49} &
  \multirow{2}{*}{4.41} &
  \multirow{2}{*}{5.48} &
  \multirow{2}{*}{4.65} &
  \multirow{2}{*}{10.49} &
  \multirow{2}{*}{4.56} \\
$\sigma=0.5$               &          &                &          &                 &          &                &          &                 \\ \midrule
$\nu =-1$ &
  \multirow{2}{*}{1.69} &
  \multirow{2}{*}{0.69} &
  \multirow{2}{*}{2.97} &
  \multirow{2}{*}{0.8} &
  \multirow{2}{*}{1.69} &
  \multirow{2}{*}{2.37} &
  \multirow{2}{*}{2.97} &
  \multirow{2}{*}{1.39} \\
$\sigma=1$                 &          &                &          &                 &          &                &          &                 \\ \bottomrule
\end{tabular}%
}
\end{table}

\begin{figure}[!t]
    \centering
    \includegraphics[width=\linewidth]{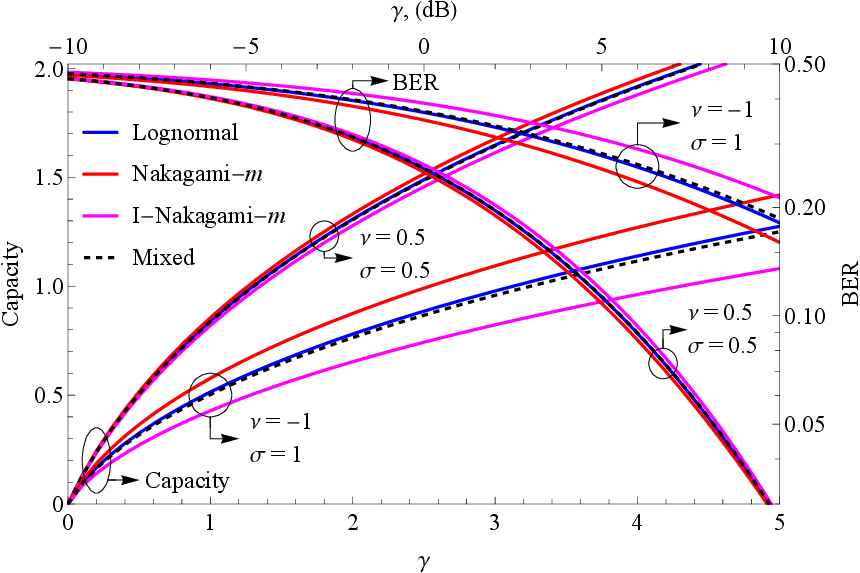}
    \caption{Lognormal Capacity and BER and its Nakagami-\emph{m}, I-Nakagami-\emph{m} and the mixed distribution equivalent.}
    \label{fig:BERCap}
\end{figure}

In Figure \ref{fig:BERCap}, the bit error ratio for an antipodal differentially coherent BPSK ($b=1$ and $a=1$) and the channel capacity are depicted. Therein, the scenario for $\nu=-1$ and $\sigma=1$ required a higher number of variates in the product. 
In this case, we set $N = 10$, which still represents a relatively small number of product terms.
On the other hand, for the other case $N = 5$ were enough to obtain a close match between the Lognormal metric and the product metric.

\begin{figure}[!t]
    \centering
    \includegraphics[width=\linewidth]{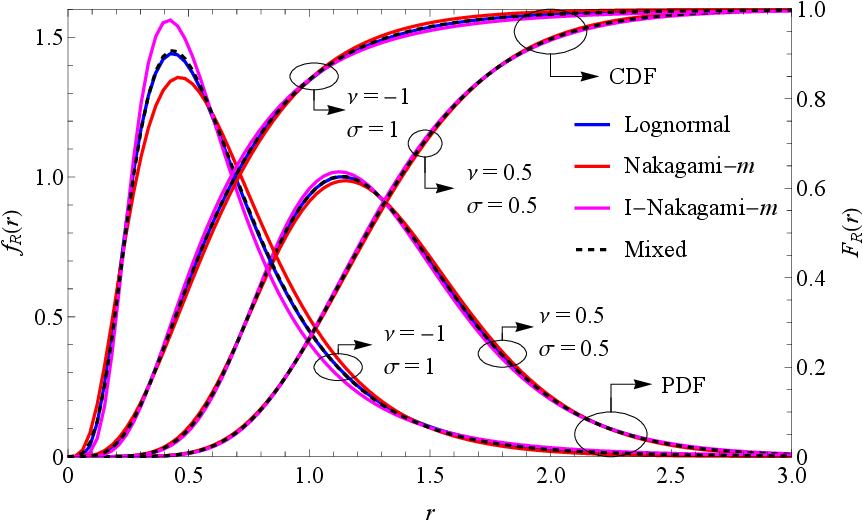}
    \caption{PDF and CDF for the $\alpha$-$\mu$ - Lognormal composite fading and the Nakagami-\emph{m}, I-Nakagami-\emph{m}, and mixed equivalents for $\alpha = 3.5$ and $\mu = 2$.}
    \label{fig:shadowV1}
\end{figure}

\begin{figure}[!t]
    \centering
    \includegraphics[width=\linewidth]{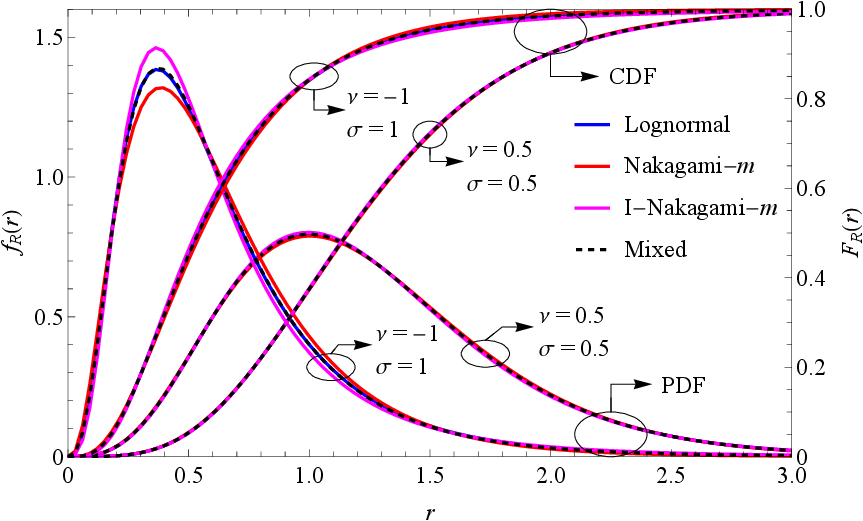}
    \caption{PDF and CDF for the Nakagami-$m$ - Lognormal composite fading and the Nakagami-\emph{m}, I-Nakagami-\emph{m}, and mixed equivalents with $m = 2$.}
    \label{fig:shadowV2}
\end{figure}

In Figure \ref{fig:shadowV1}, the PDF and CDF for the composite \am{}-Lognormal are depicted. Again, for both sets of Lognormal, the Nakagami-\emph{m} and the I-Nakagami-\emph{m} approximation was performed with $N = 5$. The chosen parameters for the \am{} part was $\alpha = 3.5$, $\mu = 2$ and $\hat{r} = 1$. In Figure \ref{fig:shadowV2}, the PDF and CDF for the Nakagami-\emph{m} -- Lognormal composite fading are depicted. The Nakagami-$m$ parameter was set as $m = 2$.

\begin{figure}[!t]
    \centering
    \includegraphics[width=\linewidth]{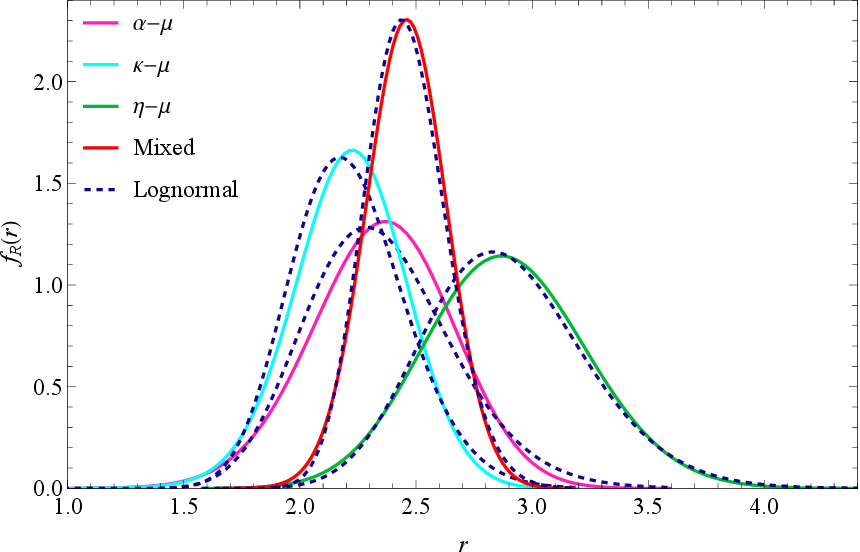}
    \caption{PDF of the product of 5 \am{}, 5 \km{}, and 5 \et{} variates and their mixed product. The dashed curves are the respective Lognormal approximation.}
    \label{fig:prod}
\end{figure}

In Figure \ref{fig:prod}, the PDF for the product of five \am{}, five \km{}, and five \et{} variates and their 
mixed product are depicted together with their respective Lognormal correspondences, all normalized by the power of 1/5 
for each set of products and by 1/15 for the mixed product. The parameters utilized are 
provided in Table \ref{tab:prod} and the Lognormal parameters were obtained individually for each product of random
variates following the procedures as described in Section \ref{sec:usefull}. The moments utilized to obtain $\sigma_Y$ was 1/5
(one over the number of variates) and for $\nu_Z$ and $\sigma_Z$ was 1/5 and 2/5. It can be observed that, even though the 
Lognormal might differ from the exact product PDF for the individual sets, the mixed scenario achieves an 
adequate fit. Of course, if the number of variates increases, each set will tend to the Lognormal exactly, so will the mixed product.

In all figures presented here, a relatively small number of terms were used in the product, and therefore a slight deviation from the corresponding exact analytical result may be observed. This perceived gap vanishes as the number of terms in the product increases.


\begin{table}[ht]
\centering
\caption{Parameters for the Variables in the Product Model}
\label{tab:prod}
\begin{tabular}{@{}ccccccc|cc@{}}
\toprule
\multicolumn{2}{c}{i}                                   & 1    & 2    & 3    & 4    & 5    & \multicolumn{2}{c}{Lognormal}                 \\ \midrule
\multirow{3}{*}{\rotatebox{90}{\am}} & \multicolumn{1}{c|}{$\alpha_i$}  & 4.45 & 4.44 & 2.97 & 4.96 & 1.40 & $\nu_X$               & $\sigma_X$            \\
                     & \multicolumn{1}{c|}{$\mu_i$}     & 1.13 & 4.22 & 2.95 & 0.52 & 4.22 & \multirow{2}{*}{4.23} & \multirow{2}{*}{0.67} \\
                     & \multicolumn{1}{c|}{$\hat{r}_i$} & 2.98 & 2.18 & 4.60 & 4.12 & 0.95 &                       &                       \\ \midrule
\multirow{3}{*}{\rotatebox{90}{\km}} & \multicolumn{1}{c|}{$\kappa_i$}  & 2.83 & 3.83 & 4.43 & 3.66 & 2.35 & $\nu_Y$               & $\sigma_Y$            \\
                     & \multicolumn{1}{c|}{$\mu_i$}     & 2.16 & 4.35 & 2.10 & 1.05 & 3.43 & \multirow{2}{*}{3.94} & \multirow{2}{*}{0.56} \\
                     & \multicolumn{1}{c|}{$\hat{r}_i$} & 2.66 & 3.66 & 4.94 & 2.04 & 0.69 &                       &                       \\ \midrule
\multirow{3}{*}{\rotatebox{90}{\et}} & \multicolumn{1}{c|}{$\eta_i$}    & 2.04 & 3.95 & 3.77 & 3.61 & 4.58 & $\nu_Z$               & $\sigma_Z$            \\
                     & \multicolumn{1}{c|}{$\mu_i$}     & 4.51 & 1.24 & 1.81 & 3.79 & 4.55 & \multirow{2}{*}{5.27} & \multirow{2}{*}{0.60} \\
                     & \multicolumn{1}{c|}{$\hat{r}_i$} & 4.50 & 4.62 & 1.61 & 2.97 & 2.76 &                       &                       \\ \bottomrule
\end{tabular}
\end{table}

\section{Final Remarks}\label{sec: Final Remarks}

This paper introduced two tractable surrogate models for the Lognormal distribution, constructed from the product of Nakagami-$m$ and I-Nakagami-$m$ variates. We derived exact formulations that enable the mapping of the product of these variates onto a Lognormal distribution and vice-versa. These models yield closed-form expressions for key performance metrics in wireless systems governed by Lognormal fading, as well as for the PDF and CDF of the composite $\alpha$-$\mu$-Lognormal fading model. A moment-matching framework was developed to map the Lognormal parameters to those of the surrogate models, enabling efficient and accurate implementation. 

To further enhance convergence, a random mixture model was proposed, which leverages the complementary characteristics of the Nakagami-$m$ and I-Nakagami-$m$ distributions. This hybrid approach markedly improves statistical fidelity and convergence rate, achieving high accuracy with substantially fewer multiplicative terms. These advantages render it particularly suited for performance analysis in practical wireless communication scenarios, where balancing analytical accuracy and computational efficiency is essential.


Additionally, the methodology was extended to accommodate heterogeneous cascaded fading channels involving $\alpha$-$\mu$, \mbox{$\kappa$-$\mu$}, and $\eta$-$\mu$ variates, demonstrating the versatility and broad applicability of the proposed framework.

Finally, beyond providing an alternative yet tractable analytical formulation, the proposed surrogate distributions offer a new perspective on how the Lognormal distribution can be analyzed and applied. They allow Lognormal-based models to be reinterpreted through equivalent product-domain representations, enabling analytical tractability, usually impaired due to the Lognormal’s inherent mathematical complexity. In this sense, the proposed framework redefines the conventional understanding of the Lognormal distribution---transforming it from a mathematically intricate model into a flexible and versatile analytical tool applicable across a wide range of scenarios.


\bibliographystyle{IEEEtran}
\bibliography{bib/bibliography}



  \vfill



\end{document}